# $L_0$-norm constraint normalized subband adaptive filtering algorithm: Performance development and AEC application

Dongxu Liu, *Student Member, IEEE*, Haiquan Zhao, *Senior Member, IEEE*, Yang Zhou

*Abstract*—Limited by fixed step-size and sparsity penalty factor, the conventional sparsity-aware normalized subband adaptive filtering (NSAF) type algorithms suffer from trade-off requirements of high filtering accurateness and quicker convergence behavior for sparse system identification. To deal with this problem, this paper proposes variable step-size $L_0$-norm constraint NSAF algorithms (VSS-$L_0$-NSAFs). We first analyze mean-square-deviation (MSD) statistics behavior of the $L_0$-NSAF innovatively in according to novel weight recursion form and arrive at corresponding expressions for the cases that background noise variance is available and unavailable, where correlation degree of system input is indicated by scaling parameter *r*. Based on derivations, we develop an effective variable step-size scheme through minimizing the upper bounds of the MSD under some reasonable assumptions and lemma. Furthermore, an effective reset strategy is incorporated into presented algorithms to tackle with non-stationary situations. Finally, numerical simulations corroborate that the proposed algorithms achieve better performance in terms of estimation accurateness and tracking capability in comparison with existing related algorithms in sparse system identification and adaptive echo cancellation circumstances.

*Index Terms*—Sparse system identification, $L_0$-NSAF algorithm, novel mean-square-deviation analysis, variable step-size.

## I. INTRODUCTION

ADAPTIVE filtering (AF) has been studied extensively for various practical requirements profiting from superior estimation accuracy and tracking capability over the past few decades [1]-[10]. In several applications, the impulse response (IR) of fundamental system being estimated exhibits sparsity characteristic commonly, meaning that only a few components of IR have significant amplitude while the rest are negligible, which is referred to as sparse system (SSI). Examples include acoustic echo cancellation (AEC) [11]-[13], channel estimation [14], acoustic feedback reduction in hearing aids [15], and underwater acoustic communications [16], to mention a few. For adaptive echo cancellation, which is main issue addressed in this paper, AF can be regarded as an efficient approach to eliminate echo channels [17]. However, the echo cancellation ability of AF depends on the design of AF algorithms (AFAs). Currently, there have reported various AFAs, and among them, the normalized-least-mean-square (NLMS) received increased attention for acoustic and speech applications owing to simplicity and robustness [18]. However, when dealing with highly correlated signals, which are usually exiting in AEC scenario (i.e., the real speech signals) [19]-[21], the NLMS demonstrates slow convergence due to large eigenvalue spread of covariance matrix of correlated signals. An effective approach to this issue is normalized subband adaptive filtering (NSAF) algorithm, which whitens system input and desired signals in multiband structure to promote the adaption [22]. Studies indicated that exploiting the *prior* sparsity information of underlying systems acquired remarkable performance improvement on AFAs, however this is not considered in above algorithms.

### A. Related works

An early outstanding work on SSI is $L_1$-norm LMS ($L_1$-LMS) algorithm benefiting from the development of compressive sensing [23], followed by several improved variants [24]-[26]. The main idea behind this technique is that through integrating $L_p$-norm regularization factor of filter weight into objection functions of the AFAs directly to quicken convergence process of non-zero elements of underlying systems (where $p = 0, 1$, namely obtaining $L_0$- and $L_1$- norms). To enhance learning performance on SSI with highly correlated input signals, the $L_1$-norm NSAF ($L_1$-NSAF) and reweighted $L_1$-NSAF ($L_1$-RNSAF) algorithms were presented by employing $L_1$- and reweighted $L_1$- norms respectively [27]. Similarly, through introducing $L_0$-norm penalty factor, the resulting $L_0$-norm constraint NSAF ($L_0$-NSAF) algorithm realized higher filtering accurateness and faster convergence [28], [29].

It is indisputable that AFAs with constant step-size generate conflicting requirements between convergence rate and filtering accuracy. To eliminate this limitation, based on NSAF, various variable step-size (VSS) variants have been established one after another. The variable step-size matrix NSAF

This work was supported in part by the National Natural Science Foundation of China under Grant 62171388, Grant 61871461, and Grant 61571374; and in part by the Fundamental Research Funds for the Central Universities under Grant 2682021ZTPY091. *(Corresponding author: Haiquan Zhao.)*

Dongxu Liu, Haiquan Zhao, and Yang Zhou are with the Key Laboratory of Magnetic Suspension Technology and Maglev Vehicle, Ministry of Education, and the School of Electrical Engineering, Southwest Jiaotong University, Chengdu, 610031, China. (e-mail: dxliu_ngu@163.com; hqzhao_swjtu@126.com; zhouyang_swjtu@126.com).

(VSSM-NSAF) was firstly designed to acquire excellent estimation accurateness, quicker convergence behavior, and outstanding tracking capability in non-stationary environments which the information of interest exhibits time-varying characteristic in comparison with NSAF [30]. To further reduce steady-state error, the variable individual step-size NSAF (VISS-NSAF) was developed through allocating individual step-size to each subband [31]. Nevertheless, though the superior learning performance was arrived at, the VISS-NSAF lost tracking ability due to monotonically decreasing property of time-variant step-size. Furthermore, several improved VSS versions have been proposed based on different optimization approaches [32]-[34]. Similarly, the sparsity-aware NSAF-type algorithms undergo restricted performance influenced by fixed step-size and sparsity penalty factor as well. To this end, through adjusting penalty factor adaptively, the adaptation-$L_1$-NSAF (A-$L_1$-NSAF) and reweighted $L_1$-NSAF (A-$L_1$-RNSAF) were presented [35]. Based on model-driven strategy, the variable-parameter- $L_1$-RNSAF (VP-$L_1$-RNSAF) and $L_0$-NSAF (VP-$L_0$-NSAF) were derived by minimizing mean-square-deviation (MSD) to jointly optimize step-size and penalty factor, with realizing good performance in terms of identification accuracy and convergence speed [36]. It is noteworthy that this algorithm supposes that background noise variance is known, however this is unavailable in most practical applications. Moreover, the above algorithms are developed under white Gaussian background noise (WGBN), to receive good convergence performance under impulsive interference environments, some robust sparsity-aware-type algorithms have also been proposed recently [37]-[40].

### B. Contributions of the Paper

In this paper, motived by the approach considered in [41], we conduct MSD performance analysis on the $L_0$-NSAF algorithm innovatively and develop an effective variable step-size scheme for performance improvement in AEC application. The main contributions of this work are denoted as follows in detail:

1) We first analyze MSD statistics behavior of the $L_0$-NSAF algorithm innovatively in according to novel weight recursion form and arrive at corresponding expressions for the cases that background noise variance is available and unavailable, where correlation degree of system inputs is indicated by scaling parameter $r$. To be specific, when noise variance is available, the MSD behavior is updated recursively through utilizing *prior* knowledge of noise variance. Otherwise, the MSD recursion is computed by resorting to exponential moving average method.

2) To solve trade-off requirements on quicker adaption ability and higher filtering accurateness, based on novel MSD derivation, a productive VSS strategy is devised with the help of some frequently-utilized assumptions and lemma, that is, propose the VSS-$L_0$-NSAF algorithms. To solve non-stationary estimation problems, a meaningful reset method is integrated into the proposed algorithms.

3) Firstly, numerical simulations are conducted to investigate the influences of different parameters related to presented algorithms on learning performance and select corresponding optimal values in detail. Secondly, through comparing with state-of-the-art algorithms, the developed algorithms demonstrate superior convergence performance in terms of estimation accuracy and tracking capability under SSI and AEC scenarios.

The key difference of optimization ideas between this work in the paper and that in [36] is as follows. Noting that our work inspects both cases where noise variance is available and unavailable based on novel MSD analysis approach, which is not considered in conventional VP-$L_0$-NSAF (as mentioned before, it is only assumed that noise variance is available). This may cause the VP-$L_0$-NSAF to become unstable in non-stationary situations. In addition, the proposed methods exhibit good adaptability to different correlated inputs through adjusting scaling parameter $r$, which is also missing in the VP-$L_0$-NSAF.

### C. Organization of the Paper

The remainder of this paper is organized as follows. Section II reviews multiband model of signals and $L_0$-NSAF algorithm. Section III carries out MSD statistics behavior analysis. Section IV derives variable step-size scheme, along with reset method integrating into proposed algorithms. Numerical simulation are provided in Section V. Section VI concludes this paper and presents future research works.

### D. Mathematics Notation

Throughout this paper, bold lowercase and uppercase letters denote vectors and matrices respectively, other notations and corresponding descriptions are provided as below.

| Notation | Description |
|---|---|
| $t$ | Time index of original sequences |
| $\tau$ | Time index of decimated sequences |
| $L$ | Length of vector of interest |
| $N$ | Number of subband |
| $(\cdot)^T$ | Transpose of vector or matrix |
| $\downarrow N$ | $N$-fold decimation |
| $E[\cdot]$ | Mathematical expectation |
| $\text{Tr}\{\cdot\}$ | Trace of matrix |
| $\|\cdot\|$ | Euclidean norm |
| $|\cdot|$ | Absolute value of scalar |
| $\text{diag}\{\cdot\}$ | Diagonal matrix |
| $\text{mod}(\cdot,\cdot)$ | Modulo operation |
| $\text{sort}(\cdot)$ | The ascending order operator |
| $\max(\cdot,\cdot)$ | Take maximum value |
| $\mathbf{I}$ | Identity matrix with compatible dimension |

## II. MODEL OF SIGNALS AND $L_0$-NSAF ALGORITHM

To begin with, the multiband signal model proposed in [22] is introduced, and then we review the $L_0$-NSAF algorithm and propose novel weight recursion form as a starting point. The performance analyses and algorithm development are carried out in subsequent sections.

## A. Multiband model of signals

Assuming that system desired output satisfies the following relation under system identification scenario:

$$d(t) = \mathbf{x}^T(t)\boldsymbol{\omega}_0 + \eta(t), \tag{1}$$

where $\boldsymbol{\omega}_0$ denotes $L$-dimensional identified unknown system, $\mathbf{x}(t)$ indicates system input, $\eta(t)$ represents zero-mean WGBN of variance $\sigma_\eta^2$.

Fig. 1 depicts the multiband-structure of SAF. Let $\boldsymbol{\omega}(\tau) = [\omega_0(\tau), \omega_1(\tau), ..., \omega_{L-1}(\tau)]^T$ be the estimation of adaptive filter of length $L$ to underlying system $\boldsymbol{\omega}_0$ at discrete time $\tau$. For highly correlated inputs, through utilizing analysis filter $H_m(z)$, the input signal $\mathbf{x}(t)$ is divided into corresponding subband signal $\mathbf{x}_m(\tau)$, and each subband input $\mathbf{x}_m(\tau)$ is approximately white Gaussian, thus speeding up the adaption process, where $m = 0, ..., N-1$. Here, the analysis filter $H_m(z)$ is described by

$$H_m(z) = \sum_{i=0}^{M-1} h_{m,i} z^{-(i-1)}, \tag{2}$$

where $h_{m,i}$ denotes $i$-th coefficient of $m$-th analysis filter.

For $m$-th subband, with critical decimation to subband desired $d_m(t)$ and output signals $y_m(t)$, the corresponding decimated results $d_{m,D}(\tau)$ and $y_{m,D}(\tau)$ are reached respectively. Then the subband output $y_{m,D}(\tau)$ is subtracted from $d_{m,D}(\tau)$ to obtain subband error signal $e_{m,D}(\tau)$ to update the coefficients $\boldsymbol{\omega}(\tau)$ of adaptive filter, so that ultimately $\boldsymbol{\omega}(\tau) = \boldsymbol{\omega}_0$, i.e., realizing underlying system identification. Furthermore, the subband error signal $e_{m,D}(\tau)$ is described as

$$\begin{aligned}e_{m,D}(\tau) &= d_{m,D}(\tau) - y_{m,D}(\tau) \\ &= \mathbf{x}_m^T(\tau)\boldsymbol{\omega}_0 + \eta_{m,D}(\tau) - \mathbf{x}_m^T(\tau)\boldsymbol{\omega}(\tau) \\ &= \mathbf{x}_m^T(\tau)\tilde{\boldsymbol{\omega}}(\tau) + \eta_{m,D}(\tau),\end{aligned} \tag{3}$$

where $\mathbf{x}_m(\tau) = [x(\tau N), x(\tau N - 1), ..., x(\tau N - L + 1)]^T$ indicates $m$-th subband input and $\eta_{m,D}(\tau)$ denotes $m$-th subbband background noise with zero-mean and variance $\sigma_{m,\eta}^2 = \sigma_\eta^2 / N$ [22]. The weight-error vector $\tilde{\boldsymbol{\omega}}(\tau)$ in (3) is defined as

$$\tilde{\boldsymbol{\omega}}(\tau) = \boldsymbol{\omega}_0 - \boldsymbol{\omega}(\tau). \tag{4}$$

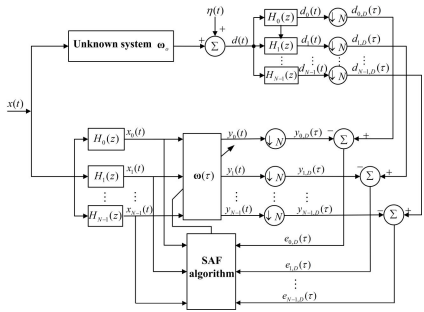

**Fig. 1.** Multiband-structure of SAF.

## B. Review of $L_0$-NSAF algorithm

First, the following optimization problem should be solved to derive the $L_0$-NSAF algorithm [28]:

$$J(\tau) = \sum_{m=0}^{N-1} \frac{e_{m,D}^2(\tau)}{\|\mathbf{x}_m(\tau)\|_2} + \rho \|\boldsymbol{\omega}(\tau)\|_0, \tag{5}$$

where $\|\boldsymbol{\omega}(\tau)\|_0$ denotes the number of non-zero components of estimation vector $\boldsymbol{\omega}(\tau)$, $\rho > 0$ indicates sparsity penalty factor for sparse system. To minimize (5), the $L_0$-norm requires to be approximated as minimization, which is considered as NP-hard issue in [24]. Through employing exponential function-based approximation scheme, the $L_0$-norm is approximated as [42]

$$\|\boldsymbol{\omega}(\tau)\|_0 \approx \sum_{j=0}^{L-1} \left(1 - e^{-\alpha|\omega_j(\tau)|}\right), \tag{6}$$

where $\alpha$ represents approximation parameter. The terms on both sides of (6) are strictly equal with parameter $\alpha$ tending to infinity. Further, (5) is re-expressed as follows by incorporating above relation:

$$J(\tau) = \sum_{m=0}^{N-1} \frac{e_{m,D}^2(\tau)}{\|\mathbf{x}_m(\tau)\|_2} + \rho \sum_{j=0}^{L-1} \left(1 - e^{-\alpha|\omega_j(\tau)|}\right), \tag{7}$$

By taking gradient of (7) with regard to $\boldsymbol{\omega}(\tau)$, and utilizing steepest descent approach and the first-order Taylor series expansions of exponential function, leading to the recursion of $L_0$-NSAF algorithm [18], [24], [28]:

$$\boldsymbol{\omega}(\tau+1) = \boldsymbol{\omega}(\tau) + \mu \sum_{m=0}^{N-1} \frac{e_{m,D}(\tau)\mathbf{x}_m(\tau)}{\|\mathbf{x}_m(\tau)\|_2} - \kappa\, \boldsymbol{f}(\tau), \tag{8}$$

where $\mu > 0$ indicates step-size, and $\kappa \triangleq \mu\rho > 0$ stands for penalty factor. $\boldsymbol{f}(\tau) = [f_0(\tau), f_1(\tau), ..., f_{L-1}(\tau)]^T$ is computed by following equation [24]:

$$f_j(\tau) = \begin{cases} -\theta^2 \omega_j(\tau) - \theta, & -1/\theta \leq w_j(\tau) < 0, \\ -\theta^2 \omega_j(\tau) + \theta, & 0 < w_j(\tau) \leq 1/\theta, \\ 0, & \text{otherwise}, \end{cases} \tag{9}$$

where $\theta > 0$ is utilized to control the range of zero-attraction, and $j = 0, ..., L-1$.

Distinguished from [28], by taking into consideration (9), (8) is re-written as another novel recursion form:

$$\boldsymbol{\omega}(\tau+1) = \boldsymbol{\omega}(\tau) + \mu \sum_{m=0}^{N-1} \frac{e_{m,D}(\tau)\mathbf{x}_m(\tau)}{\|\mathbf{x}_m(\tau)\|_2} - \kappa\left(\mathbf{S}(\tau)\boldsymbol{\omega}(\tau) + \mathbf{G}(\tau)\right), \tag{10}$$

where $\mathbf{S}(\tau)$ is a diagonal matrix of size $L \times L$, and $\mathbf{G}(\tau)$ is column vector of size $L \times 1$, which are defined as follows:

$$\mathbf{S}(\tau) = \text{diag}\{s_0(\omega_0(\tau)), s_1(\omega_1(\tau)), ..., s_{L-1}(\omega_{L-1}(\tau))\}, \tag{11}$$

$$\mathbf{G}(\tau) = \left(g_0(\omega_0(\tau)), g_1(\omega_1(\tau)), ..., g_{L-1}(\omega_{L-1}(\tau))\right)^T, \tag{12}$$

where $s_j(\omega_j(\tau))$ and $g_j(\omega_j(\tau))$ are provided as

$$s_j(\omega_j(\tau)) = \begin{cases} -\theta^2, & 0 < |\omega_j(\tau)| \leq 1/\theta, \\ 0, & \text{otherwise}, \end{cases} \quad (13)$$

$$g_j(\omega_j(\tau)) = \begin{cases} -\theta, & -1/\theta \leq \omega_j(\tau) < 0, \\ \theta, & 0 < \omega_j(\tau) \leq 1/\theta, \\ 0, & \text{otherwise}. \end{cases} \quad (14)$$

### III. MEAN SQUARE DEVIATION ANALYSES

In this section, the MSD statistics behavior of the $L_0$-NSAF algorithm for the cases that background noise variance is available and unavailable is firstly analyzed as basic, and then deriving the optimal step-size.

For mathematical tractability, the following assumptions are beneficial:

*Assumption 1:* The subband signals $\mathbf{x}_m(\tau)$, $\eta_{m,D}(\tau)$, and $\tilde{\boldsymbol{\omega}}(\tau)$ are mutually independent [19].

*Assumption 2:* The signals $\eta_{m,D}(\tau)$ and $\boldsymbol{\omega}(\tau)$ are also independent [28], [29].

#### A. Case where background noise variance is available

Due to the observability of input signals, for given set $X_\tau \triangleq \{\mathbf{x}_i \mid 0 \leq i < \tau\}$, the conditional covariance matrix of $\tilde{\boldsymbol{\omega}}(\tau)$ is defined by

$$\mathbf{P}(\tau) \triangleq E\left[\tilde{\boldsymbol{\omega}}(\tau)\tilde{\boldsymbol{\omega}}^{\mathrm{T}}(\tau) \mid X_\tau\right], \quad (15)$$

and the MSD of the algorithm is determined as

$$p(\tau) \triangleq E\left[\tilde{\boldsymbol{\omega}}^{\mathrm{T}}(\tau)\tilde{\boldsymbol{\omega}}(\tau) \mid X_\tau\right] = \mathrm{Tr}\left[\mathbf{P}(\tau)\right]. \quad (16)$$

Under (3), (4), and (10), the recursion of $\tilde{\boldsymbol{\omega}}(\tau)$ is established as follows:

$$\begin{aligned}
&\tilde{\boldsymbol{\omega}}(\tau+1) \\
&= \tilde{\boldsymbol{\omega}}(\tau) - \mu \sum_{m=0}^{N-1} \frac{\mathbf{x}_m(\tau) e_{m,D}(\tau)}{\|\mathbf{x}_m(\tau)\|^2} + \kappa(\mathbf{S}(\tau)\boldsymbol{\omega}(\tau) + \mathbf{G}(\tau)) \\
&= \tilde{\boldsymbol{\omega}}(\tau) - \mu \sum_{m=0}^{N-1} \frac{\mathbf{x}_m(\tau)\left(\mathbf{x}_m^{\mathrm{T}}(\tau)\tilde{\boldsymbol{\omega}}(\tau) + \eta_{m,D}(\tau)\right)}{\|\mathbf{x}_m(\tau)\|^2} + \kappa(\mathbf{S}(\tau)\boldsymbol{\omega}(\tau) + \mathbf{G}(\tau)) \\
&= \mathbf{F}(\tau)\tilde{\boldsymbol{\omega}}(\tau) - \mu \sum_{m=0}^{N-1} \frac{\mathbf{x}_m(\tau)\eta_{m,D}(\tau)}{\|\mathbf{x}_m(\tau)\|^2} + \kappa(\mathbf{S}(\tau)\boldsymbol{\omega}(\tau) + \mathbf{G}(\tau)),
\end{aligned} \quad (17)$$

where $\mathbf{F}(\tau) = \left(\mathbf{I} - \mu \sum_{m=0}^{N-1} \mathbf{x}_m(\tau)\mathbf{x}_m^{\mathrm{T}}(\tau)/\|\mathbf{x}_m(\tau)\|^2\right)$ denotes transition matrix.

Post-multiplying (17) by its transpose, taking expectation and considering (15), one can arrive at

$$\begin{aligned}
&\mathbf{P}(\tau+1) \\
&= \mathbf{F}(\tau)\mathbf{P}(\tau)\mathbf{F}^{\mathrm{T}}(\tau) - \mu\left(\mathbf{P}_1(\tau) + \mathbf{P}_1^{\mathrm{T}}(\tau)\right) - \mu\kappa\left(\mathbf{P}_2(\tau) + \mathbf{P}_2^{\mathrm{T}}(\tau)\right) \\
&\quad + \kappa\left(\mathbf{P}_3(\tau) + \mathbf{P}_3^{\mathrm{T}}(\tau)\right) + \mu^2 \mathbf{P}_4(\tau) + \kappa^2 \mathbf{P}_5(\tau),
\end{aligned} \quad (18)$$

with

$$\mathbf{P}_1(\tau) = \sum_{m=0}^{N-1} E\left[\frac{\mathbf{F}(\tau)\tilde{\boldsymbol{\omega}}(\tau)\eta_{m,D}(\tau)\mathbf{x}_m^{\mathrm{T}}(\tau)}{\|\mathbf{x}_m(\tau)\|^2} \mid X_\tau\right], \quad (19)$$

$$\mathbf{P}_2(\tau) = \sum_{m=0}^{N-1} E\left[\frac{\mathbf{x}_m(\tau)\eta_{m,D}(\tau)}{\|\mathbf{x}_m(\tau)\|^2}(\mathbf{S}(\tau)\boldsymbol{\omega}(\tau) + \mathbf{G}(\tau))^{\mathrm{T}} \mid X_\tau\right], \quad (20)$$

$$\mathbf{P}_3(\tau) = E\left[\mathbf{F}(\tau)\tilde{\boldsymbol{\omega}}(\tau)(\mathbf{S}(\tau)\boldsymbol{\omega}(\tau) + \mathbf{G}(\tau))^{\mathrm{T}} \mid X_\tau\right], \quad (21)$$

$$\mathbf{P}_4(\tau) = \sum_{m=0}^{N-1} E\left[\frac{\eta_{m,D}^2(\tau)\mathbf{x}_m(\tau)\mathbf{x}_m^{\mathrm{T}}(\tau)}{\left[\|\mathbf{x}_m(\tau)\|^2\right]^2} \mid X_\tau\right], \quad (22)$$

$$\mathbf{P}_5(\tau) = E\left[(\mathbf{S}(\tau)\boldsymbol{\omega}(\tau) + \mathbf{G}(\tau))(\mathbf{S}(\tau)\boldsymbol{\omega}(\tau) + \mathbf{G}(\tau))^{\mathrm{T}} \mid X_\tau\right]. \quad (23)$$

Based on *assumptions 1-2*, one can reach that

$$\mathbf{P}_1(\tau) = \sum_{m=0}^{N-1} \mathbf{F}(\tau) E\left[\tilde{\boldsymbol{\omega}}(\tau)\eta_{m,D}(\tau) \mid X_\tau\right] \frac{\mathbf{x}_m^{\mathrm{T}}(\tau)}{\|\mathbf{x}_m(\tau)\|^2} = 0, \quad (24)$$

$$\begin{aligned}
&\mathbf{P}_2(\tau) \\
&= \sum_{m=0}^{N-1} E\left[\frac{\mathbf{x}_m(\tau)\eta_{m,D}(\tau)}{\|\mathbf{x}_m(\tau)\|^2}(\mathbf{S}(\tau)\boldsymbol{\omega}(\tau) + \mathbf{G}(\tau))^{\mathrm{T}} \mid X_\tau\right] \\
&= \sum_{m=0}^{N-1} E\left[\frac{\mathbf{x}_m(\tau)\eta_{m,D}(\tau)}{\|\mathbf{x}_m(\tau)\|^2}\boldsymbol{\omega}^{\mathrm{T}}(\tau)\mathbf{S}(\tau) + \frac{\mathbf{x}_m(\tau)\eta_{m,D}(\tau)}{\|\mathbf{x}_m(\tau)\|^2}\mathbf{G}^{\mathrm{T}}(\tau) \mid X_\tau\right] \\
&= 0.
\end{aligned} \quad (25)$$

Introducing orthogonality principle [43] and unbiased estimation $E\left[e_{m,D}(\tau) \mid X_\tau\right] = 0$ [44] lead to following result:

$$\begin{aligned}
&\mathbf{P}_3(\tau) \\
&= E\left[\mathbf{F}(\tau)\tilde{\boldsymbol{\omega}}(\tau)(\mathbf{S}(\tau)\boldsymbol{\omega}(\tau) + \mathbf{G}(\tau))^{\mathrm{T}} \mid X_\tau\right] \\
&= \mathbf{F}(\tau)E\left[\tilde{\boldsymbol{\omega}}(\tau)\boldsymbol{\omega}^{\mathrm{T}}(\tau) \mid X_\tau\right]\mathbf{S}(\tau) + \mathbf{F}(\tau)E\left[\tilde{\boldsymbol{\omega}}(\tau) \mid X_\tau\right]\mathbf{G}^{\mathrm{T}}(\tau) \\
&= 0.
\end{aligned} \quad (26)$$

Substituting (24)-(26) into (18) can obtain that

$$\mathbf{P}(\tau+1) = \mathbf{F}(\tau)\mathbf{P}(\tau)\mathbf{F}^{\mathrm{T}}(\tau) + \mu^2 \mathbf{P}_4(\tau) + \kappa^2 \mathbf{P}_5(\tau). \quad (27)$$

Taking trace on both sides of (27) and employing cyclic property of trace, one can yield

$$p(\tau+1) = \mathrm{Tr}\left\{\mathbf{F}^{\mathrm{T}}(\tau)\mathbf{F}(\tau)\mathbf{P}(\tau)\right\} + \mu^2 \mathrm{Tr}\left\{\mathbf{P}_4(\tau)\right\} + \kappa^2 \mathrm{Tr}\left\{\mathbf{P}_5(\tau)\right\}. \quad (28)$$

Under the definition of $\mathbf{F}(\tau)$, the first term of (28) is calculated by

$$\mathrm{Tr}\left\{\mathbf{F}^{\mathrm{T}}(\tau)\mathbf{F}(\tau)\mathbf{P}(\tau)\right\} = \mathrm{Tr}\left\{\mathbf{P}(\tau)\right\} + (\mu^2 - 2\mu)\mathrm{Tr}\left\{\sum_{m=0}^{N-1}\mathbf{A}_m(\tau)\mathbf{P}(\tau)\right\}, \quad (29)$$

where $\mathbf{A}_m(\tau) = \mathbf{x}_m(\tau)\mathbf{x}_m^{\mathrm{T}}(\tau)/\|\mathbf{x}_m(\tau)\|^2$.

Considering following trace inequality [45] in which has been successfully employed to analyze MSD behavior of the algorithms [46]-[48]:

$$\sum_{j=1}^{L}\lambda_j(U)\lambda_{L-j+1}(V) \leq \mathrm{Tr}\{UV\} \leq \sum_{j=1}^{L}\lambda_j(U)\lambda_j(V), \quad (30)$$

where $U$ and $V$ are Hermitian positive semidefinite matrices of size $L\times L$, and $\lambda_j(U)$ is the $j$-th largest eigenvalue of the matrix $U$.

Through incorporating this property into (29), the following inequality is achieved:

$$\mathrm{Tr}\{\mathbf{F}^\mathrm{T}(\tau)\mathbf{F}(\tau)\mathbf{P}(\tau)\} \leq \mathrm{Tr}\{\mathbf{P}(\tau)\} + (\mu^2-2\mu)N\lambda_{\min}(\mathbf{P}(\tau)), \quad (31)$$

where $\lambda_{\min}(\mathbf{P}(\tau))$ is minimum eigenvalue of matrix $\mathbf{P}(\tau)$. By considering that inequality $L\lambda_{\min}(\mathbf{P}(\tau)) \leq \mathrm{Tr}\{\mathbf{P}(\tau)\}$ always holds, therefore the following approximation is established with existing a constant $r_1 \geq 1$:

$$\lambda_{\min}(\mathbf{P}(\tau)) \approx \frac{\mathrm{Tr}\{\mathbf{P}(\tau)\}}{r_1 L}. \quad (32)$$

where $r_1$ stands for scaling parameter in which incarnates the degree of correlation of input signals. In particular, when system input is WGN, the input covariance matrix is well-conditioned and demonstrates small eigenvalue spread. Under this case, the parameter $r_1$ is chosen as 1. For correlated input signals, the input covariance matrix becomes ill-conditioned and corresponding value should be appropriately determined.

By combining this approximation, (31) can be re-written as

$$\mathrm{Tr}\{\mathbf{F}^\mathrm{T}(\tau)\mathbf{F}(\tau)\mathbf{P}(\tau)\} \leq p(\tau) + (\mu^2-2\mu)\frac{Np(\tau)}{r_1 L}. \quad (33)$$

Adopting *assumption 1* again results in

$$\mathrm{Tr}\{\mathbf{P}_4(\tau)\} = \sum_{m=0}^{N-1}\frac{\sigma_{m,\eta}^2}{\|\mathbf{x}_m(\tau)\|^2}. \quad (34)$$

For the calculation of $\mathrm{Tr}\{\mathbf{P}_5(\tau)\}$ in (28), it is determined as follows:

$$E\left[(\mathbf{S}(\tau)\boldsymbol{\omega}(\tau)+\mathbf{G}(\tau))^\mathrm{T}(\mathbf{S}(\tau)\boldsymbol{\omega}(\tau)+\mathbf{G}(\tau))\mid X_\tau\right] \approx \alpha(\tau), \quad (35a)$$

$$\alpha(\tau) = \gamma\alpha(\tau-1)+(1-\gamma)\left[(\mathbf{S}(\tau)\boldsymbol{\omega}(\tau)+\mathbf{G}(\tau))^\mathrm{T}(\mathbf{S}(\tau)\boldsymbol{\omega}(\tau)+\mathbf{G}(\tau))\right], \quad (35b)$$

where exponential moving average approach is utilized to as approximation for conditional expectation term, and $\gamma$ indicates forgetting factor with $0<\gamma<1$ [41].

Finally, combining (33)-(35) into (28), the upper bound of MSD is denoted recursively as:

$$p(\tau+1) \leq p(\tau) + (\mu^2-2\mu)\frac{Np(\tau)}{r_1 L} + \mu^2\sum_{m=0}^{N-1}\frac{\sigma_{m,\eta}^2}{\|\mathbf{x}_m(\tau)\|^2} + \kappa^2\alpha(\tau), \quad (36)$$

For step-size bounded in $0<\mu<2$ of the NSAF [22], the convergence behavior of the $L_0$-NSAF is warranted when the following relationship is satisfied:

$$\left|1+(\mu^2-2\mu)\frac{N}{r_1 L}\right| < 1. \quad (37)$$

It is obvious that this inequality is always established, which indicates that the stabilization of the $L_0$-NSAF is guaranteed at the adaption process when noise variance is available (in AEC, $N \ll L$ is common for NSAF-type algorithms due to the long acoustic echo channel).

### B. Case where background noise variance is unavailable

Given that the knowledge on background noise variance is unavailable in most practical applications, instead of utilizing known information of noise variance $\sigma_\eta^2$, the MSD expression is described recursively in terms of another form.

By combining (4), (10) along with (15), one can realize that $\mathbf{P}(\tau+1)$

$$= \mathbf{P}(\tau) - \mu\left(\mathbf{Q}_1(\tau)+\mathbf{Q}_1^\mathrm{T}(\tau)\right) - \mu\kappa\left(\mathbf{Q}_2(\tau)+\mathbf{Q}_2^\mathrm{T}(\tau)\right) \quad (38)$$
$$-\kappa\left(\mathbf{Q}_3(\tau)+\mathbf{Q}_3^\mathrm{T}(\tau)\right)+\mu^2\mathbf{Q}_4(\tau)+\kappa^2\mathbf{Q}_5(\tau),$$

with

$$\mathbf{Q}_1(\tau) = \sum_{m=0}^{N-1}E\left[\frac{\tilde{\boldsymbol{\omega}}(\tau)e_{m,D}(\tau)\mathbf{x}_m^\mathrm{T}(\tau)}{\|\mathbf{x}_m(\tau)\|^2}\tilde{\boldsymbol{\omega}}^\mathrm{T}(\tau)\mid X_\tau\right], \quad (39)$$

$$\mathbf{Q}_2(\tau) = \sum_{m=0}^{N-1}E\left[\frac{\mathbf{x}_m(\tau)e_{m,D}(\tau)}{\|\mathbf{x}_m(\tau)\|^2}(\mathbf{S}(\tau)\boldsymbol{\omega}(\tau)+\mathbf{G}(\tau))^\mathrm{T}\mid X_\tau\right], \quad (40)$$

$$\mathbf{Q}_3(\tau) = E\left[\tilde{\boldsymbol{\omega}}(\tau)(\mathbf{S}(\tau)\boldsymbol{\omega}(\tau)+\mathbf{G}(\tau))^\mathrm{T}\mid X_\tau\right], \quad (41)$$

$$\mathbf{Q}_4(\tau) = \sum_{m=0}^{N-1}E\left[\frac{e_{m,D}^2(\tau)\mathbf{x}_m(\tau)\mathbf{x}_m^\mathrm{T}(\tau)}{\left[\|\mathbf{x}_m(\tau)\|^2\right]^2}\mid X_\tau\right], \quad (42)$$

$$\mathbf{Q}_5(\tau) = E\left[(\mathbf{S}(\tau)\boldsymbol{\omega}(\tau)+\mathbf{G}(\tau))(\mathbf{S}(\tau)\boldsymbol{\omega}(\tau)+\mathbf{G}(\tau))^\mathrm{T}\mid X_\tau\right]. \quad (43)$$

Substituting (3) into (39) results in
$$\mathbf{Q}_1(\tau)$$
$$= \sum_{m=0}^{N-1}E\left[\frac{\tilde{\boldsymbol{\omega}}(\tau)\left(\tilde{\boldsymbol{\omega}}^\mathrm{T}(\tau)\mathbf{x}_m(\tau)+\eta_{m,D}(\tau)\right)\mathbf{x}_m^\mathrm{T}(\tau)}{\|\mathbf{x}_m(\tau)\|^2}\mid X_\tau\right]$$

$$= \sum_{m=0}^{N-1}E\left[\frac{\tilde{\boldsymbol{\omega}}(\tau)\tilde{\boldsymbol{\omega}}^\mathrm{T}(\tau)\mathbf{x}_m(\tau)\mathbf{x}_m^\mathrm{T}(\tau)}{\|\mathbf{x}_m(\tau)\|^2}\mid X_\tau\right], \quad (44)$$

By applying orthogonality principle and *assumption 2* again, the following results are obtained:
$$\mathbf{Q}_2(\tau)$$
$$= \sum_{m=0}^{N-1}E\left[\frac{\mathbf{x}_m(\tau)\left(\mathbf{x}_m^\mathrm{T}(\tau)\tilde{\boldsymbol{\omega}}(\tau)+\eta_{m,D}(\tau)\right)}{\|\mathbf{x}_m(\tau)\|^2}(\mathbf{S}(\tau)\boldsymbol{\omega}(\tau)+\mathbf{G}(\tau))^\mathrm{T}\mid X_\tau\right]$$
$$= 0, \quad (45)$$

$$\mathbf{Q}_3(\tau) = E\left[\tilde{\boldsymbol{\omega}}(\tau)\left(\mathbf{S}(\tau)\boldsymbol{\omega}(\tau)+\mathbf{G}(\tau)\right)^\mathrm{T} \mid X_\tau\right] = 0. \quad (46)$$

Substituting (44)-(46) into (38) arrives at that

$$\mathbf{P}(\tau+1) = \mathbf{P}(\tau) - \mu\left(\mathbf{Q}_1(\tau)+\mathbf{Q}_1^\mathrm{T}(\tau)\right) + \mu^2 \mathbf{Q}_4(\tau) + \kappa^2 \mathbf{Q}_5(\tau). \quad (47)$$

Taking trace on both sides of (47), we have

$$p(\tau+1) = p(\tau) - 2\mu \operatorname{Tr}\left\{\mathbf{Q}_1^\mathrm{T}(\tau)\right\} + \mu^2 \operatorname{Tr}\left\{\mathbf{Q}_4(\tau)\right\} + \kappa^2 \operatorname{Tr}\left\{\mathbf{Q}_5(\tau)\right\}. \quad (48)$$

Following similar way in (35), $\operatorname{Tr}\{\mathbf{Q}_4(\tau)\}$ is calculated by

$$E\left[\frac{e_{m,D}^2(\tau)}{\|\mathbf{x}_m(\tau)\|^2} \mid X_\tau\right] \approx \upsilon_m(\tau), \quad (49)$$

$$\upsilon_m(\tau) = \gamma \upsilon_m(\tau-1) + (1-\gamma)\left(e_{m,D}^2(\tau)/\|\mathbf{x}_m(\tau)\|^2\right). \quad (50)$$

Utilizing trace inequality (30) again, the upper bound of the MSD is indicated by

$$p(\tau+1) \le p(\tau) - \frac{2\mu N}{r_2 L} p(\tau) + \mu^2 \sum_{m=0}^{N-1} \upsilon_m(\tau) + \kappa^2 \alpha(\tau). \quad (51)$$

where $r_2 \ge 1$ is also scaling parameter that characterizes the system input correlation.

Likewise, the following relation holds always:

$$\left|1 - \frac{2\mu N}{r_2 L}\right| < 1, \quad (52)$$

which implies the stability of the $L_0$-NSAF algorithm is assured during the iterations when noise variance is unavailable.

## IV. Development of Algorithm

The step-size affects learning performance of AFAs remarkably. An inappropriate step-size value leads to a significant decrease in algorithm's performance or even makes the algorithm fail to converge to steady-state. However, for the most of exiting algorithms, the step-size is usually chosen by trial and error, which may not be optimal in theory. To this end, this paper proposes a novel variable step-size approach through minimizing the upper bounds of MSD of the algorithm for both situations that background noise variance is available and unavailable to achieve performance improvement in terms of filtering accurateness and adaption speed. Moreover, the non-stationary situations, such as underlying systems vary abruptly, are also considered cautiously to further enhance the tracking capability of the proposed algorithms.

### A. Derivation of variable step-size approach

Firstly, the step-size $\mu$ is replaced by its time-varying form $\mu_0(\tau)$ in (10), (36), and (51). For deriving optimal step-size value, the following optimal solution should be resolved:

$$\mu(\tau) = \arg\min_{\mu} p(\tau+1). \quad (53)$$

Then the following two cases are examined respectively:

*Case 1:* When noise variance $\sigma_\eta^2$ is available, the upper bound of MSD (36) can be re-expressed as

$$p(\tau+1) \le p(\tau) + \Gamma_1(\tau), \quad (54a)$$

$$\Gamma_1(\tau) \triangleq \left(\mu(\tau)^2 - 2\mu(\tau)\right)\frac{Np(\tau)}{r_1 L} + \mu(\tau)^2 \sum_{m=0}^{N-1} \frac{\sigma_{m,\eta}^2}{\|\mathbf{x}_m(\tau)\|^2} + \kappa^2 \alpha(\tau). \quad (54b)$$

Taking derivative of (54) with respect to $\mu(\tau)$ and then let it to be zero, one arrives at

$$\mu(\tau) = \frac{Np(\tau)}{Np(\tau) + r_1 L\left(\sum_{m=0}^{N-1} \sigma_{m,\eta}^2 / \|\mathbf{x}_m(\tau)\|^2 + \rho^2 \alpha(\tau)\right)}. \quad (55)$$

*Case 2:* When noise variance $\sigma_\eta^2$ is not available, one can obtain that

$$p(\tau+1) \le p(\tau) + \Gamma_2(\tau), \quad (56a)$$

$$\Gamma_2(\tau) \triangleq \left(\mu(\tau)^2 - 2\mu(\tau)\right)\frac{Np(\tau)}{r_2 L} + \mu(\tau)^2 \sum_{m=0}^{N-1} \frac{\sigma_{m,\eta}^2}{\|\mathbf{x}_m(\tau)\|^2} + \kappa^2 \alpha(\tau). \quad (56b)$$

Following similar approach, we have

$$\mu(\tau) = \frac{Np(\tau)}{r_2 L \left(\sum_{m=0}^{N-1} \upsilon_m(\tau) + \rho^2 \alpha(\tau)\right)}. \quad (57)$$

### B. Practical considerations

For algorithm implementation, the following issues should be considered:

1) *Non-negativity guarantee:* In accordance with analysis results provided in (36) and (51), it is evident that the MSD value $p(\tau)$ is positive for *cases 1* and *2*. It should be mentioned that due to some approximations and assumptions have been introduced to derive the MSD. However, this may result in negative values of $p(\tau)$, thereby producing significant oscillations or negative values on step-size to effect the stability of the algorithms. For this purpose, the recursions (54) and (56) for $p(\tau)$ are modified as follows:

$$\bar{p}(\tau+1) = \begin{cases} p(\tau) + \left(\mu(\tau)^2 - 2\mu(\tau)\right)\frac{Np(\tau)}{r_1 L} \\ + \mu^2 \sum_{m=0}^{N-1} \frac{\sigma_{m,\eta}^2}{\|\mathbf{x}_m(\tau)\|^2} + \kappa^2 \alpha(\tau), \text{ Case 1,} \\ p(\tau) - \frac{2\mu(\tau)N}{r_2 L} p(\tau) \\ + \mu(\tau)^2 \sum_{m=0}^{N-1} \upsilon_m(\tau) + \kappa^2 \alpha(\tau), \text{ Case 2,} \end{cases} \quad (58a)$$

$$p(\tau+1) = \max(0, \bar{p}(\tau+1)), \quad (58b)$$

where $\bar{p}(\tau)$ is upper bound of $p(\tau)$ at iteration $\tau$.

Then introducing (58) into (54) and (56), the step-size $\mu(\tau)$ is always positive at all iterations, which arrives at non-negativity guarantee.

2) *Non-stationary situation:* For AEC scenario, the estimated echo channel is commonly time-varying, which belongs to non-stationary situations. In this case, the proposed algorithms acquire obvious degradation in terms of tracking performance (this has been proved in simulation experiments, which is not shown in the paper). To overcome this shortcoming, a productive reset algorithm is integrated into the algorithms to reach superior tracking capability [49].

As described in Table II, $V_T$ and $V_D$ refer to positive integers with $V_T > V_D$, $\mathbf{M} = \text{diag}(1,...,1,0,...,0)$ indicates diagonal matrix with first $V_T - V_D$ elements being one. $\phi > 0$ represents threshold parameter to detect environmental changes. When the system varies abruptly, the initializations of filter weight $\boldsymbol{\omega}(\tau)$ and MSD $p(\tau)$ are conducted to adapt the change of environment. Otherwise, the algorithms are updated normally. Typical values of $V_T$, $V_D$, $\varepsilon$, and $\phi$ are set as follows: $V_T = 3L$, $V_D = 0.75V_T$, $\varepsilon = 10^{-6}$, and $\phi = 10^{-3}$ [49].

**Reset mechanism:**
if $\mod(\tau, V_T/N) = 0$

$$\mathbf{R} = \text{sort}\left(\frac{|e(\tau)|}{\|\mathbf{x}(\tau)\|_2 + \varepsilon},...,\frac{|e(\tau - V_T + 1)|}{\|\mathbf{x}(\tau - V_T + 1)\|_2 + \varepsilon}\right)^{\mathrm{T}}$$

$$z_n = \mathbf{R} \mathbf{M} \mathbf{R}^{\mathrm{T}} / (V_T - V_D)$$

end if

$\Delta_\tau = (z_n - z_o)/\sqrt{\mu(\tau)}$

if $\Delta_\tau > \phi$

$\boldsymbol{\omega}(\tau) = \mathbf{0}$; $\alpha(\tau) = 0$; $\upsilon_m(\tau) = 0$; $p(\tau) = 1$

else

Weight update

end

$z_o = z_n$

*3) Numerical difficulty:* Owing to the existence of silent periods of speech signals in AEC, the algorithms suffer from numerical difficulty in adaption process. Therefore, a regularization factor $\delta > 0$ is incorporated to avoid this dilemma, and the developed algorithms are represented as:

$$\boldsymbol{\omega}(\tau+1) = \boldsymbol{\omega}(\tau) + \mu(\tau)\sum_{m=0}^{N-1}\frac{e_{m,D}(\tau)\mathbf{x}_m(\tau)}{\|\mathbf{x}_m(\tau)\|_2 + \delta} - \kappa\left(\mathbf{S}(\tau)\boldsymbol{\omega}(\tau) + \mathbf{G}(\tau)\right). \quad (59)$$

## V. NUMERICAL SIMULATIONS

Numerical simulations are executed to investigate learning behaviors of the algorithms under sparse system identification and adaptive echo cancellation scenarios in MATLAB. The system to be identified with sparse characteristic is originated by normal distribution with zero-mean and unit variance, and the positions of non-zero elements are selected randomly, unless otherwise stated. The tap-length of the adaptive filter is presumed to be the same as that of estimated unknown system. For all simulations, the filter coefficients are initialized with zero-vector. For system input, various correlated signals are utilized which generated by first-order autoregressive (AR(1)) and second-order autoregressive (AR(2)) random processes according to

$$x(t) = 0.95x(t-1) + z(t), \quad (60)$$
$$x(t) = -0.1x(t-1) - 0.8x(t-2) + z(t), \quad (61)$$

where $z(t)$ indicates zero-mean WGN with unit variance. And the real speech signals with non-stationary characteristic are employed for AEC application. Furthermore, two or four cosine-modulated filter banks with lengths $M = 17$ and 33 respectively are selected to divide system input and desired signals in SAF structure. The system out background noise is modeled by WGN with signal-to-noise ratio (SNR) of 30dB or 20dB, which is defined as follows:

$$\text{SNR} = 10\log_{10}[E(y^2(t)]/[E(\eta^2(t)], \quad (62)$$

where $y(t) = \mathbf{x}^{\mathrm{T}}(t)\boldsymbol{\omega}_0$ is system out in time domain.

The normalized misalignment is exploited as performance measure, which is defined by:

$$10\log_{10}(\|\boldsymbol{\omega}_0 - \boldsymbol{\omega}(\tau)\|^2 / \|\boldsymbol{\omega}_0\|^2). \quad (63)$$

Finally, ensemble averaging was executed over 100 Monte-Carlo experimental results to depict simulation curves, unless otherwise stated.

### A. Perspectives on the proposed algorithms

In this section, we examine the influences of scaling parameter $r_1/r_2$, sparsity penalty factor $\rho$, and the number of subband $N$ on presented algorithms (55) and (57) with AR(1) and AR(2) inputs. The sparse IR with length $L = 100$ and 4 non-zero components is employed as identified system, which is demonstrated in Fig. 2. In all simulations, $\delta = 0.01$ is chosen to prevent numerical difficulty. Likewise, this is also suitable for simulations in *Section B*.

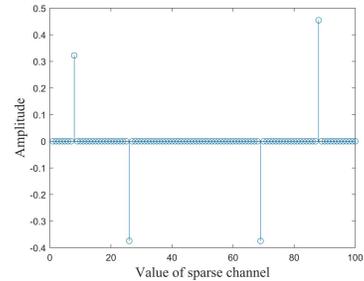

Fig. 2. Sparse IR with length $L = 100$ and 4 non-zero components.

TABLE I
PARAMETER SETTINGS OF ALL CONSIDERING ALGORITHMS

| Algorithms | $L_0$-NSAF | VSSM-NSAF | VISS-NSAF | NVSS-NSAF | VP-$L_0$-NSAF | Proposed (eq. (55)) | Proposed (eq. (57)) |
|---|---|---|---|---|---|---|---|
| Parameters | $\mu/\rho/\beta$ | $\alpha/\varepsilon_1/\varepsilon_2$ | $\text{Tr}\{P(0)\}/\beta$ | $\alpha/\lambda$ | $\eta/\lambda/\mu_{\max}/\beta$ | $\gamma/\rho/\theta/r_1$ | $\gamma/\rho/\theta/r_2$ |
| Fig. 7(a) | 0.17/1e-5/5 | 0.99/10⁻³/10⁻⁵ | 1/4.5 | 0.994/6 | 0.98/0.95/1/5 | 0.99/4e-5/5/1.4 | 0.99/1e-4/5/1.8 |
| Fig. 7(b) | 0.1/1e-5/5 | 0.99/10⁻³/10⁻⁵ | 1/4.5 | 0.994/6 | 0.98/0.95/1/5 | 0.99/4e-4/5/1 | 0.99/4e-4/5/1.4 |
| Fig. 8(a) | 0.17/1e-5/5 | 0.93/10⁻³/10⁻⁵ | 1/4.5 | 0.994/6 | 0.99/0.97/1/5 | 0.99/1e-4/5/1.4 | 0.99/1e-4/5/2.5 |
| Fig. 8(b) | 0.1/1e-5/5 | 0.93/10⁻³/10⁻⁵ | 1/4.5 | 0.994/6 | 0.99/0.97/1/5 | 0.99/1e-4/5/1 | 0.99/1e-4/5/1.8 |
| Fig. 9(a) | 0.17/1e-5/5 | 0.93/10⁻³/10⁻⁵ | 1/4.5 | 0.994/6 | 0.98/0.95/1/5 | 0.992/4e-5/5/1.4 | 0.99/1e-4/5/1.8 |
| Fig. 9(b) | 0.1/1e-5/5 | 0.93/10⁻³/10⁻⁵ | 1/4.5 | 0.994/6 | 0.98/0.95/1/5 | 0.992/4e-4/5/1 | 0.99/4e-4/5/1.4 |
| Fig. 10(a) | 0.17/1e-5/5 | 0.93/10⁻³/10⁻⁵ | 1/4.5 | 0.994/6 | 0.98/0.95/0.35/5 | 0.992/1e-4/5/1.4 | 0.99/1e-5/5/2.5 |
| Fig. 10(b) | 0.17/1e-5/5 | 0.93/10⁻³/10⁻⁵ | 1/4.5 | 0.994/6 | 0.98/0.95/1/5 | 0.992/4e-4/5/1 | 0.98/1e-4/5/1.8 |
| Fig. 14 | 0.35/1e-6/2 | 0.9997/10⁻⁶/10⁻⁸ | 100/4.5 | 0.999/3 | 0.999/0.999/1/5 | 0.85/1e-6/2/4 | 0.96/1e-6/2/11 |
| Fig. 15 | 0.35/1e-6/2 | 0.9997/10⁻⁶/10⁻⁸ | 100/4.5 | 0.999/3 | 0.999/0.999/1/5 | 0.85/1e-6/2/4 | 0.96/1e-6/2/11 |

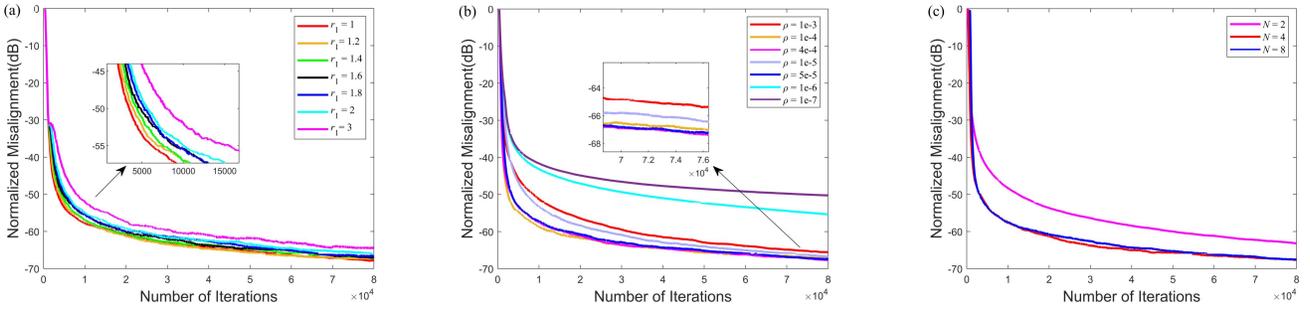

**Fig. 3.** Normalized misalignment learning curves of (55) under AR(1) correlated input. (a) Effect of $r_1$; (b) effect of $\rho$; (c) effect of $N$.

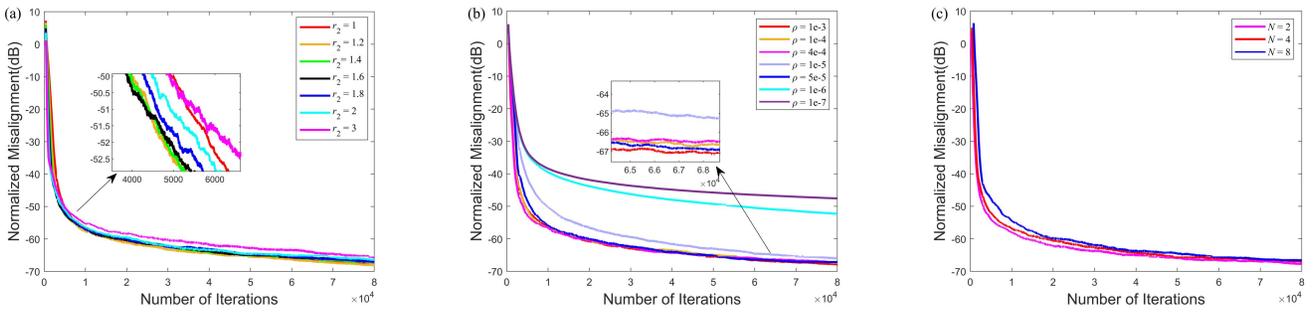

**Fig. 4.** Normalized misalignment learning curves of (57) under AR(1) correlated input. (a) Effect of $r_2$; (b) effect of $\rho$; (c) effect of $N$.

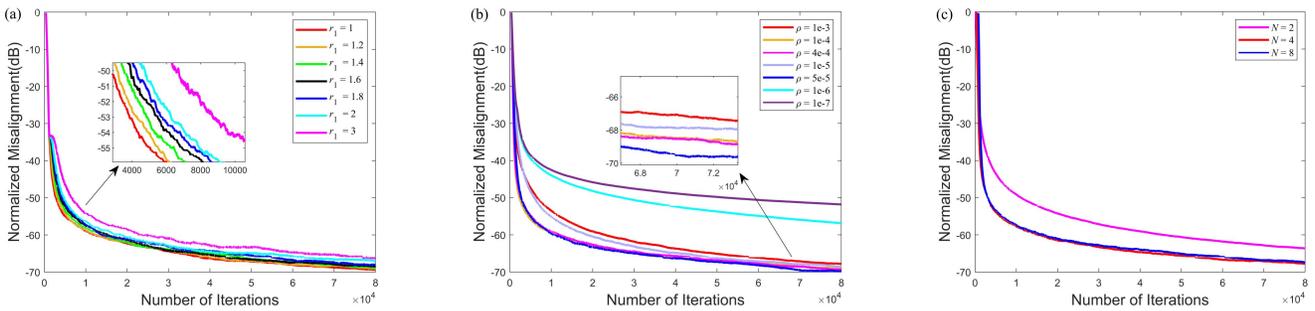

**Fig. 5.** Normalized misalignment learning curves of (55) under AR(2) correlated input. (a) Effect of $r_1$; (b) effect of $\rho$; (c) effect of $N$.

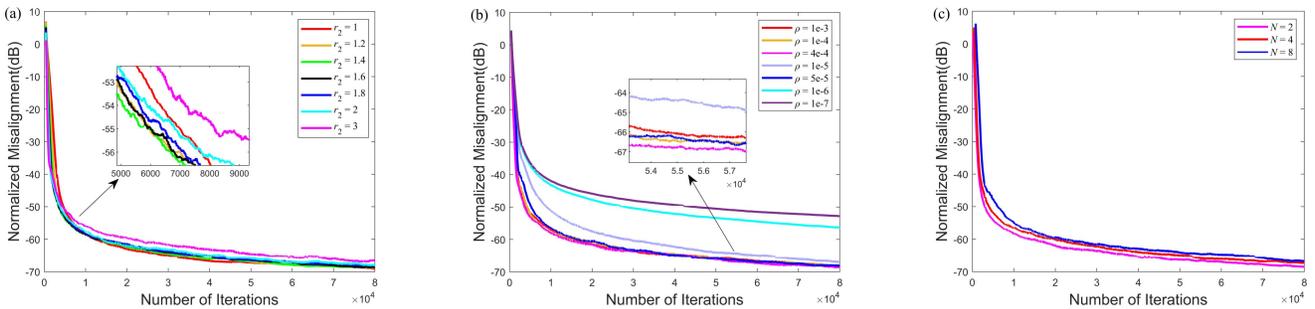

**Fig. 6.** Normalized misalignment learning curves of (57) under AR(2) correlated input. (a) Effect of $r_2$; (b) effect of $\rho$; (c) effect of $N$.

Figs. 3-4 depict learning behavior curves for (55) and (57) with different parameter settings under AR(1) input. From Fig. 3, we have following observations: 1) the scaling parameter $r_1 \in [1, 1.2]$ provides good performance in terms of original adaption speed and steady-state misalignment in comparison with other values; 2) the influence of penalty factor $\rho$ is significant on learning performance of (55) with different selections. In particular, $\rho \in [1e-4, 4e-4, 5e-5]$ demonstrates higher filtering accuracy under similar original adaption. This inspires us that an appropriate value of $\rho$ is essential for realizing superior performance on underlying system identification with different sparsity levels; 3) the convergence behaviors exhibit some variation through utilizing different numbers of subband $N$. Generally, the larger number of subband leads to higher computational complexity. It is evident that $N = 4$ is ideal choice to obtain excellent performance without more complexity increased by contrasting with other values. Furthermore, similar conclusions are made in Fig. 4 for (57), and we are not to discuss here. It is worth mentioning that, unlike (55), when $N$ is selected as 2 instead of 4, (57) can achieve comparable convergence to (55) with lower computational burden.

When system input is AR(2) correlated signal, it can be seen that the proposed two algorithms (55) and (57) still reach equivalent adaption behavior in comparison with AR(1) input, which is displayed in Figs. 5-6. The reason behind this may be that by employing the SAF structure, these two algorithms obtain whitening subband signals with similar property under these two different correlated input signals.

### B. Performance evaluation with conventional algorithms

In this section we investigate the learning performance of the proposed two algorithms in comparison with $L_0$-NSAF [28], VSSM-NSAF [30], VISS-NSAF [31], new variable step-size NSAF (NVSS-NSAF) [32], and VP-$L_0$-NSAF [36] algorithms. The estimated system is the same as in *Section A*. To examine behavior changes in non-stationary environments, the IR is converted to $-\mathbf{\omega}_0$ in the middle of adaptive process. For fair comparisons, the parameters are set so that all considered algorithms arrive at optimal performance under similar original convergence, which are summarized in Table I.

Figs. 7-8 describe learning behavior curves of the algorithms with various system inputs. In accordance with Figs. 7-8, we have following observations: 1) due to employ the *prior* sparsity property of underlying system, the sparsity-aware NSAF-type algorithms including $L_0$-NSAF, VP-$L_0$-NSAF, and proposed VSS-$L_0$-NSAFs realize higher filtering accuracy by contrasting with other algorithms through taking the advantage of $L_0$-norm constraint strategy under AR(1) input; 2) compared to conventional $L_0$-NSAF, profiting from VSS strategy, the presented algorithms (55) and (57) acquire obviously improved performance in terms of estimation accurateness, which corroborates the effectiveness of devised approach; 3) when noise variance is available, (57) results in comparable or slightly lower steady-state error than (55), with all maintaining highest filtering accuracy and acceptable tracking capability in competing algorithms under various cases.

Figs. 9-10 demonstrate learning curves of the comparing algorithms with background noise variance varied. As seen that the compared algorithms lead to similar performance to the simulation results in Figs. 7-8. However, when the SNR is changed unexpectedly, all considering algorithms receive different degrees of performance degradation. In particular, owing to the unknowability of noise variance, (55) acquires higher steady-state estimation error. On the contrary, it is interesting that (57) does not utilize this information to calculate optimal step-size for estimating underlying system, thus retaining relatively good steady-state estimation performance after the SNR changes of all the algorithms.

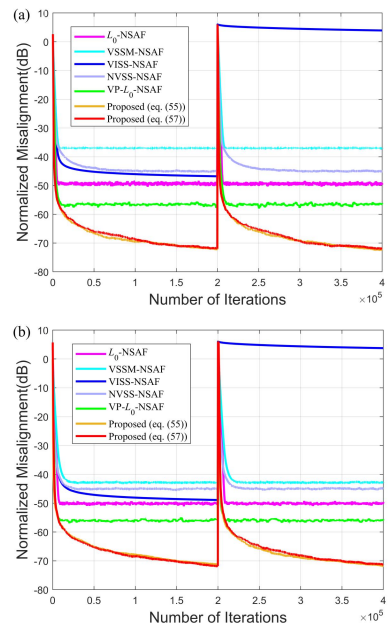

**Fig. 7.** Normalized misalignment learning curves of all algorithms with underlying system changes abruptly under AR(1) input. (a) $N = 2$; (b) $N = 4$.

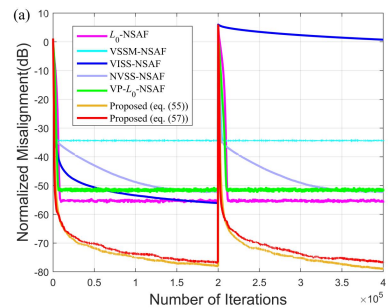

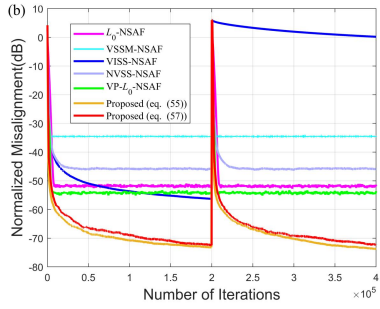

**Fig. 8.** Normalized misalignment learning curves of all algorithms with underlying system changes abruptly under AR(2) input. (a) $N = 2$; (b) $N = 4$.

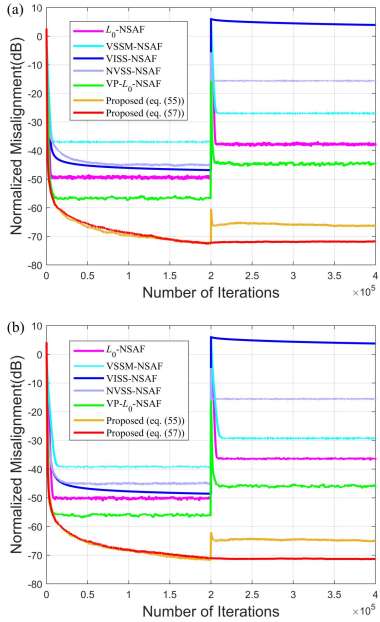

**Fig. 9.** Normalized misalignment learning curves of all algorithms with background noise variance varied from 30dB to 20dB under AR(1) input. (a) $N = 2$; (b) $N = 4$.

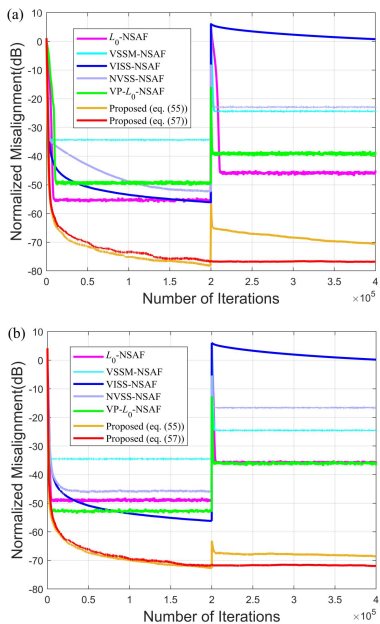

**Fig. 10.** Normalized misalignment learning curves of all algorithms with background noise variance varied from 30dB to 20dB under AR(2) input. (a) $N = 2$; (b) $N = 4$.

### C. Adaptive echo cancellation scenario

Next we consider a practical AEC application to evaluate the proposed algorithms with schematic diagram shown in Fig. 11 [41]. A network echo path from G168 Recommendation is as IR, which is illustrated in Fig. 12 [50]. Because of time-varying characteristic of echo path, it is varied by multiplying -1 at the middle of input samples. The real speech signals including far- and near- signals are depicted in Fig. 13. The near-end signal is introduced to the AEC system at $1.5 \times 10^5 \sim 3 \times 10^5$ iterations to model double-talk scenario. Parameter settings are exhibited in Table I, which refer to the suggestions in the literature. Additionally, $\delta = \sigma_x^2$ is introduced to the algorithms considering silent periods of speech signals except for the VSSM-NSAF ($\delta = 15\sigma_x^2$ is chosen), where $\sigma_x^2$ indicates the power of system input. Simulation results are obtained through one experiment.

Fig. 14 illustrates learning curves of the algorithms in AEC with speech inputs. As observed that: 1) the VP-$L_0$-NSAF reaches quicker adaption process compared to the $L_0$-NSAF through using VP approach, while receiving higher estimation error at steady-state stage; 2) by taking advantage of the VSS and reset algorithm, the proposed algorithms (55) and (57) exhibit higher identification accurateness, quicker convergence behavior, and acceptable tracking performance after system changes in comparison with other algorithms such as VSSM-NSAF, VISS-NSAF, and NVSS-NSAF; 3) (55) acquires slightly improvement in terms of steady-state filtering accurateness than (57) when noise variance is available whether underlying system changes or not.

When double-talk scenario occurs, one can notice that all algorithms (except for (57)) tend to diverge to varying degrees as shown in Fig. 15. It should be stated that the reason for performance degradation of (55) is that noise variance changes after near-end signal is included in system output, thus producing wrong estimation information to underlying system. However, benefiting from moving average method (49), (57) realizes excellent robustness to the change of background noise variance. Therefore, (57) may be more suitable choice in AEC application to eliminate or reduce echo channel.

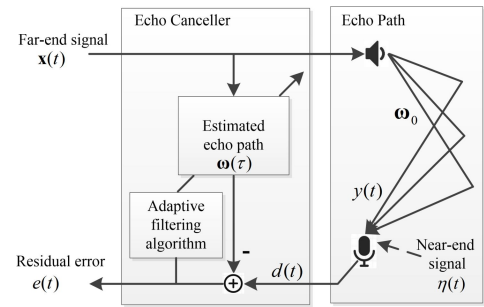

**Fig. 11.** Schematic diagram of AEC.

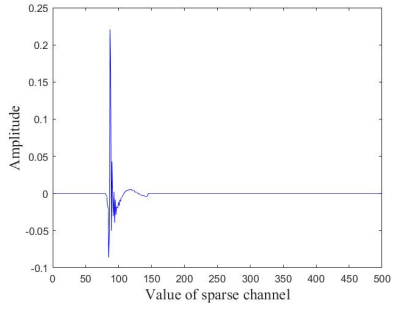

**Fig. 12.** Network echo channel in AEC.

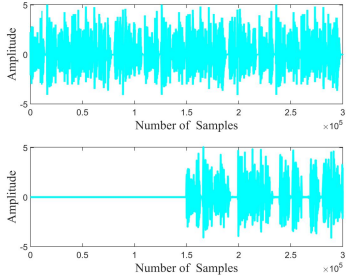

**Fig. 13.** Speech signals in AEC. Far-end speech signal (top); near-end speech signal (bottom).

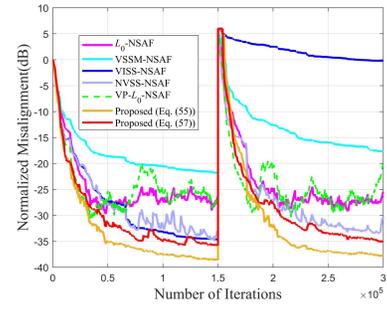

**Fig. 14.** Normalized misalignment learning curves of all algorithms with underlying system changes suddenly in AEC.

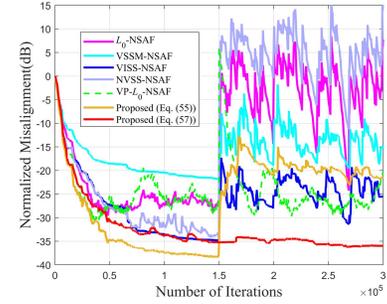

**Fig. 15.** Normalized misalignment learning curves of all algorithms under double-talk situation in AEC.

## VI. CONCLUSION AND FUTURE WORKS

This paper proposed the VSS-$L_0$-NSAF algorithms to achieve performance improvement in terms of adaption speed and filtering accurateness. The MSD statistics behavior of the $L_0$-NSAF was firstly derived based on novel analyses method, with reflecting the correlated characteristic of input signals through scaling parameter $r$. Corresponding MSD expressions for both background noise variance was available and unavailable have been obtained. Based on MSD behavior, a novel variable step-size approach was designed by minimizing the upper bounds of the MSD. By considering practical implementation, a constructive reset mechanism was incorporated into the algorithms to strengthen robustness in non-stationary environments. Numerical simulations under SSI and AEC circumstances supported that the proposed algorithms realized superior convergence behavior in terms of estimation accuracy and tracking capability in comparison with other related algorithms, which verified the validity of designed strategies.

In future works, the proposed method may be further extended to other sparsity-aware-type AFAs such as sparsity-aware- affine projection algorithm [51] and recursive-least-square algorithm [52], which are two meaningful approaches to accelerate adaption process for correlated input signals. Furthermore, the developed algorithms were implemented in real-valued domain. Research on AFAs in complex-valued domain, which is required in practical applications such as direction-of-arrival estimation [53] and transform domain signal processing [54], it is possible to be applied to widely linear complex-valued AFAs including augmented complex-valued NSAF [55], widely linear complex-valued affine projection algorithm [56], and augmented complex-valued estimation-input NSAF [57], however this is next research direction.